\documentclass[runningheads]{llncs}
\usepackage[T1]{fontenc}
\usepackage{cite}
\usepackage{amsmath,amssymb,amsfonts}
\usepackage{adjustbox}
\usepackage{mfirstuc-english}
\usepackage{algorithmic}
\usepackage{graphicx}
\usepackage{textcomp}
\usepackage{xcolor}
\usepackage{xurl}
\usepackage{makecell}
\def\BibTeX{{\rm B\kern-.05em{\sc i\kern-.025em b}\kern-.08em
    T\kern-.1667em\lower.7ex\hbox{E}\kern-.125emX}}
\begin{document}

\title{An Empirical Evaluation of BMSSP and Dijkstra's Algorithm for the Lightning Network}

\author{Danila Valko\inst{1,2}\orcidID{0000-0002-8058-7539} \and
Rohan Paranjpe\inst{3} \and
Jorge {Marx G\'omez}\inst{1}\orcidID{0000-0002-7833-7549}}
\authorrunning{D. Valko et al.}

\institute{
Carl von Ossietzky Universität Oldenburg, Germany
\and OFFIS – Institute for Information Technology, Germany 
\and Independent Researcher, USA
\\
\email{danila.valko@offis.de}
}

\maketitle

\begin{abstract}
Efficient routing is critical for payment channel networks (PCNs) such as the Lightning Network (LN), where shortest-path computations are commonly performed using Dijkstra-based algorithms. Recent theoretical work introduced the Bounded Multi-Source Shortest Path (BMSSP) algorithm, which achieves an asymptotically faster running time than Dijkstra's algorithm on sparse directed graphs. This paper presents an empirical evaluation of BMSSP on real-world LN topology snapshots and compares its performance with a standard Dijkstra implementation written in Rust. Across five LN snapshots and 100 randomized source-node trials per snapshot, BMSSP consistently exhibited higher runtimes than Dijkstra's algorithm, with statistically significant differences observed in all evaluated datasets. These findings indicate that, for current LN graph sizes and the implementations studied, BMSSP does not provide practical runtime advantages despite its stronger asymptotic guarantees. Rather than demonstrating a routing acceleration technique, this work contributes an LN-specific empirical case study of a recently proposed shortest-path algorithm and highlights the gap between theoretical complexity improvements and practical performance on contemporary payment-channel-network topologies.
\end{abstract}


\section{Introduction}
Payment channel networks (PCNs) are a prominent solution to the scalability limitations of blockchain systems. A PCN allows users to conduct multiple off-chain transactions by opening and maintaining bidirectional payment channels, thus avoiding the cost and latency of recording every transaction on the base blockchain. At the core of PCN performance is the routing problem: given a network of payment channels, a sender must efficiently discover a viable payment path that satisfies capacity and fee requirements. The efficiency of this routing directly affects transaction latency, success rate, and user experience.

The Lightning Network (LN) is the most widely used PCN, built as a second-layer protocol on top of Bitcoin. It has become both a practical payment infrastructure and a research testbed for future PCNs\cite{Zabka2024, Rohrer2019}. Given its scale and adoption, even small improvements in routing efficiency can have a significant impact on its usability, scalability and performance\cite{Valko2024}.

Currently, most LN implementations rely on variants of Dijkstra's/Yen's algorithm to compute shortest paths over the entire payment channel network represented as a graph. These implementations use standard adjacency-list representations and binary or Fibonacci heaps to achieve runtimes of $O(|E| + |V|~log~|V|)$, where $V$ is the set of nodes and $E$ is a set of channels in the PCN graph $G=(V,~E)$. While Dijkstra's algorithm has long been considered optimal in practice, recent theoretical work has demonstrated otherwise. In particular, Duan et al.\cite{Duan2025arXiv} have recently introduced the Bounded Multi-Source Shortest Path (BMSSP) algorithm, which theoretically achieves a faster time complexity of $O(|E|~log^{2/3}~|V|)$ on sparse directed graphs, potentially outperforming Dijkstra's algorithm in this setting\cite{Duan2025}.

This empirical paper investigates how BMSSP behaves on real Lightning Network topologies and whether its theoretical asymptotic advantages translate into practical performance gains at current LN scales.

Specifically, we extend an LN-related codebase with an implementation of BMSSP on \textit{Rust} and evaluate its runtime performance against the standard implementation of Dijkstra's algorithm on real LN topology data.
\section{Background}

\subsection{The BMSSP Algorithm}

Duan et al.'s BMSSP algorithm\cite{Duan2025, Duan2025arXiv} introduces the first proven deterministic improvement over Dijkstra's approach for the single-source shortest path problem\cite{Thorup} on directed graphs with non-negative real weights. As mentioned above, BMSSP achieves a runtime complexity of $O(|E|~log^{2/3}~|V|)$, which is strictly faster than Dijkstra's algorithm in the sparse regime. This breakthrough lies in circumventing Dijkstra's $O(|V|~log~|V|)$ sorting barrier: while Dijkstra's keeps a sorted list of all nodes via a priority-queue, BMSSP recursively partitions the graph and applies a small number of Bellman-Ford-style relaxations\cite{Duan2025} to keep the frontier relatively small and reduce computational overhead.

BMSSP provides an asymptotic speedup in sparse graphs, precisely the kinds of graphs LN graphs fall into. Although its requirements are minimal (non-negative edge weights), implementing BMSSP is non-trivial, and demands careful management of recursion, distance partitions, and deterministic update schedules within the comparison-addition model. Here we refer interested readers to the original paper\cite{Duan2025, Duan2025arXiv} and a detailed technical explanation in \cite{Paranjpe2025}.

\section{Pathfinding Task in the Lightning Network}
The LN is naturally sparse. Although the LN can contain thousands of nodes and tens of thousands of payment channels, the average degree per node is low, roughly on the order of 8. For example, Zabka et al.\cite{Zabka2024} reported that the LN graph exhibits a skewed degree distribution, with only a few hubs maintaining hundreds of channels while most nodes connect to fewer than ten peers. This sparsity is critical in evaluating algorithms such as BMSSP, which are specifically optimized for sparse graphs.

Routing in LN clients is currently based on conventional variants of Dijkstra's single-source algorithm, improved with priority queues. For example, the \textit{LND}\cite{Saraswathi2025} client uses a binary heap-based Dijkstra's variant, while some other prototypes experiment with Fibonacci heaps to achieve $O(|E| + |V|~log~|V|)$ time complexity. In practice, this translates into acceptable runtimes for current LN sizes, but routing performance remains a bottleneck as the network grows.

The LN graph further satisfies the requirements for BMSSP:
\begin{itemize}
\item It is directed, since payment fees and liquidity constraints introduce directionality in channel usability.
\item Edge weights are real, non-negative values, reflecting fees, base costs, and channel capacities.
\end{itemize}

These properties position LN as a generally suitable testbed for evaluating BMSSP performance in more realistic conditions.

\section{Method}

\subsection{BMSSP Implementation}

As of yet, no official reference implementation of BMSSP has been released by its authors. To bridge this gap, we contacted the authors, who confirmed the absence of a codebase. However, an initial \textit{Rust} prototype was recently developed by Rohan Paranjpe and shared in a technical blog post 25 August, 2025\cite{Paranjpe2025}. This prototype was tested on road-network graphs, a canonical example of sparse graphs, and did not show runtime improvements consistent with the theoretical guarantees. However, the evaluation was limited to this specific domain and did not address PCNs or LN-like graphs. Thus, we decided to further extend this implementation for the experimental evaluation within the LN framework.

\subsection{Benchmarks and Evaluation Strategy}
First, to evaluate BMSSP in a realistic LN setting, we curated a set of five network snapshots from publicly available LN topology data\cite{datapaper2025} (see also gossip protocol snapshots in \cite{lngossip}). Each representative snapshot was selected from the midpoint of the year, covering the available period from 2019 to 2023 (Tab. \ref{tab01}).

\begin{table}[!ht]
    \centering
    \caption{\capitalisewords{Network Snapshot Characteristics}}\label{tab01}
    \begin{adjustbox}{width=0.8\textwidth}
    \begin{tabular}{ccccc}
    \hline
        \textit{Snapshot Date} & \textit{Nodes} & \textit{Channels} & \makecell{\textit{Avg. Degree} \\± \textit{Std. Dev.}} & \textit{Degree Entropy} \\ \hline
        2019-06-17 & 4287 & 30652 & 14.30 ± 52.23 & 0.60 \\ 
        2020-06-15 & 5697 & 28743 & 10.09 ± 41.19 & 0.50 \\ 
        2021-06-15 & 10343 & 42746 & 8.27 ± 42.47 & 0.44 \\ 
        2022-06-05 & 15648 & 78468 & 10.03 ± 51.56 & 0.47 \\ 
        2023-07-16 & 15071 & 64196 & 8.52 ± 46.44 & 0.43 \\ \hline
    \end{tabular}
    \end{adjustbox}
    ~\\ 
    \vspace{0.2em}
    \small{\raggedright Note. The degree distribution entropy is normalized.\par}
\end{table}

The raw topology data were cleaned and converted into a weighted graph, where weights corresponded to routing fees and channel characteristics reflecting native pathfinding heuristic for \textit{LND} software client\cite{KumbleAnonimity}. 

In the LN, actual edge weights dynamically depend on both the payment amount and a fuzzy factor. The latter can vary stochastically based on previous channel failures, risk parameters, and other conditions that might be defined differently across software clients. Since our focus is on algorithmic performance with respect to network topology rather than payment dynamics, we did not model the fuzzy factor. Moreover, payment amounts were assumed to be equal and set to the average payment size. 

To satisfy BMSSP's assumption of unique shortest-path distances, we applied a deterministic tie-breaking transformation to edge weights. Specifically, a small perturbation was added to each edge weight based on a fixed ordering of edges. The perturbation magnitude was chosen to be several orders of magnitude smaller than the original routing costs, ensuring that shortest-path rankings were preserved whenever paths differed in total cost. The same transformed graph was supplied to both BMSSP and Dijkstra. Consequently, this preprocessing step affected both algorithms identically and does not favour either method in the runtime comparison.

Second, for each topology snapshot, we generated a fixed set of 100 source nodes using a predetermined random seed. The same source nodes were evaluated by both algorithms, yielding paired observations for each snapshot. Runtime differences were therefore attributable to algorithmic behaviour rather than variation in source-node selection.

Third, for each run and compared algorithm, we measured the total runtime for completing a single-source shortest paths problem from a chosen source node. Because identical source nodes were evaluated by both algorithms, the experimental design is naturally paired. We therefore analysed runtime differences using paired statistical comparisons (Wilcoxon signed-rank tests) for each network snapshot. 

Before benchmarking, we verified that BMSSP and Dijkstra produced identical shortest-path distance values on each LN snapshot for a common set of source nodes. For each tested source node, we compared the distance assigned to every reachable vertex by both algorithms. No significant discrepancies were observed. Where multiple shortest paths with identical total weight existed, path representations were allowed to differ provided that the computed shortest-path distance remained identical. This validation step ensured that runtime comparisons were performed between functionally equivalent shortest-path computations.

\subsection{Real-world Considerations}

While the structural properties of the LN make it a natural candidate for shortest path algorithms, real-world routing in LN is significantly more complex than the abstract formulation considered in classical graph theory. In practice, pathfinding must operate under multiple dynamic constraints that extend beyond static edge weights.

For instance, modern LN clients increasingly explore multi-path payment (MPP) strategies\cite{mpp1, mpp2}, where a transaction is split across multiple disjoint paths to improve success probability and reduce fees. MPP fundamentally changes the routing problem from a single shortest path computation to a more complex flow optimization problem. Algorithms like BMSSP and Dijkstra's are designed for single-source shortest paths and do not directly account for such extensions. Therefore, while our evaluation provides insights into core pathfinding performance, it does not capture the full spectrum of routing strategies deployed in practice.

The LN also evolves continuously, with nodes and channels being added, removed, or updated over time. This introduces the challenge of dynamic graph updates, where routing algorithms must operate on frequently changing topologies. In such settings, recomputing shortest paths from scratch for each query may become inefficient, motivating the need for incremental or adaptive algorithms. While both Dijkstra's algorithm and BMSSP are introduced and evaluated here in a static context, their behaviour under dynamic conditions remains an open question and an important direction for future research.

Despite these complexities, the simplified model adopted in this work remains a useful abstraction for isolating algorithmic performance. By focusing on static graphs with deterministic weights, we are able to directly compare the computational efficiency of BMSSP and Dijkstra without confounding factors introduced by network dynamics or probabilistic routing. This allows us to evaluate whether theoretical improvements in shortest path computation translate into measurable benefits in a controlled yet realistic approximation of LN topology.

\section{Results}
As outlined in the method section, we evaluated the competitor algorithms by performing 100 independent runs for each method on LN topology snapshots, using fixed seeds for randomly varying the source nodes. 

These experiments were run on a 2.6 GHz 6-core Intel i7 processor with 16 GB of RAM on a Macbook Pro running Sequoia 15.4.1. In each run, we recorded the total runtime required to solve the single-source shortest path problem from the chosen source node.

\begin{figure*}[t]
\includegraphics[scale=0.5]{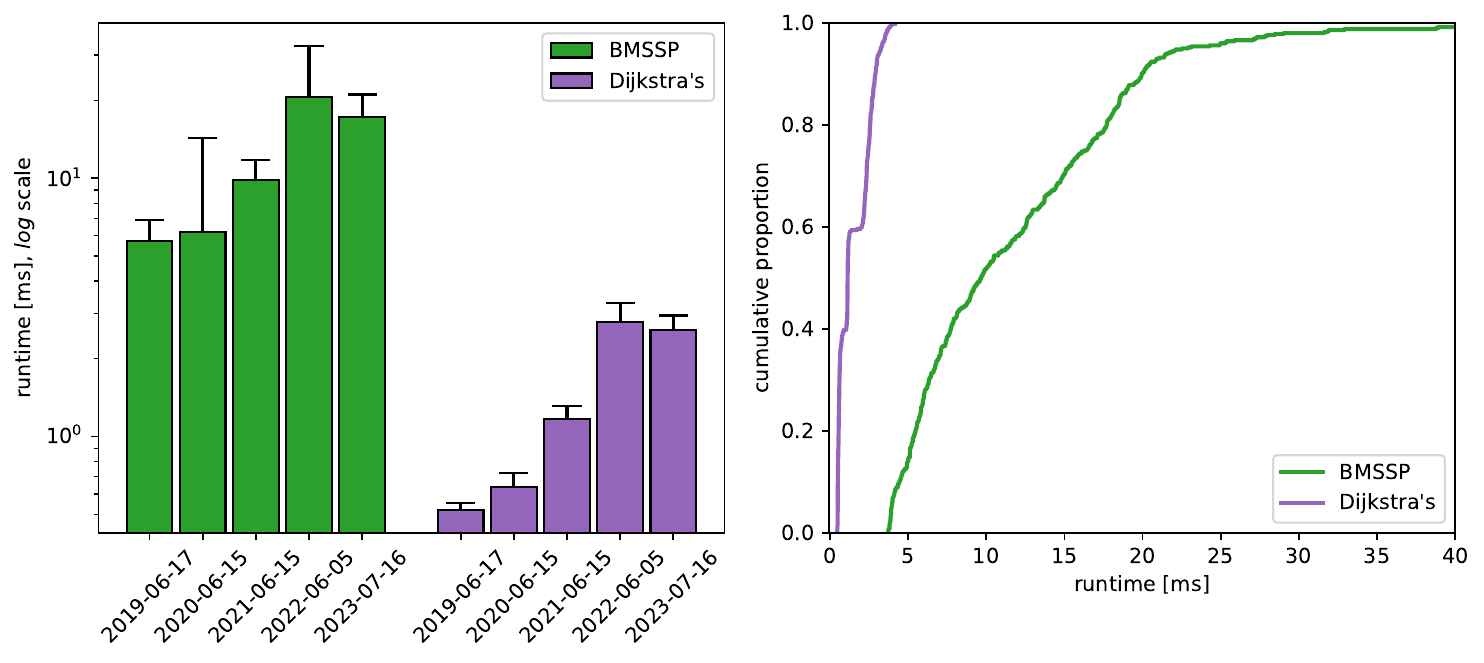}
\vspace{-0.5em}
\caption{\capitalisewords{Runtime statistics across network snapshots and their cumulative distributions}}\label{fig01}
    \vspace{0.3em}
    \small{\raggedright Note. Bars indicate the mean values for each dated topology snapshot, with whiskers showing the standard deviation. The empirical cumulative distribution is computed for each competing algorithm over all experimental runs across snapshots.\par}
\end{figure*}

The left panel of Fig.~\ref{fig01} presents the average runtime statistics per snapshot. Each bar indicates the mean runtime over 100 runs for a given snapshot, labelled by its corresponding date. As shown, in the better case the runtime of BMSSP (green bars) is nearly twice as high as that of the standard Dijkstra's implementation (purple bars).

To rigorously assess performance differences, we compared cumulative runtime distributions across all runs (the right panel of Fig.~\ref{fig01}) and applied the \textit{Mann–Whitney U} test. The results confirm a highly significant difference, with $U = 249890.0$ and $p < 0.0001$.

\begin{figure}[t]
\centering
\includegraphics[scale=0.7]{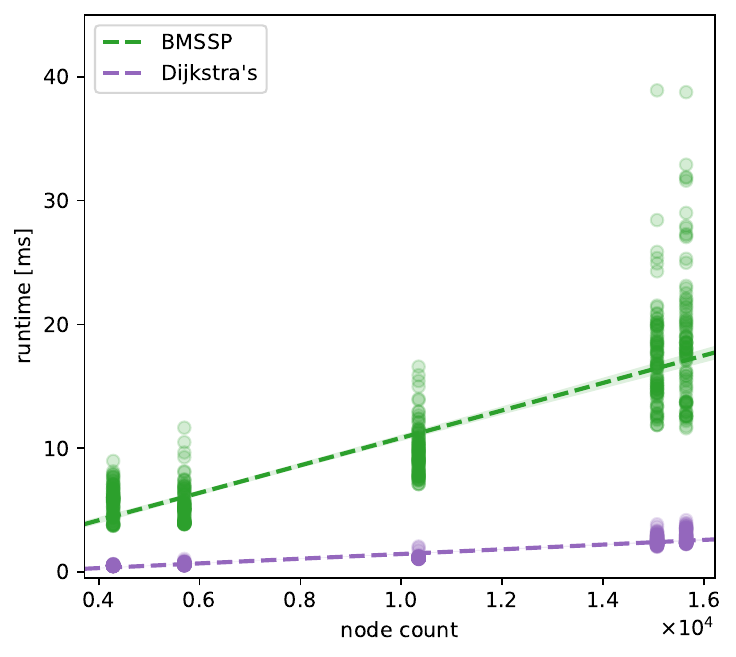}
\vspace{-0.5em}
\caption{\capitalisewords{Scaling of runtime with network size}}\label{fig02}
    \vspace{0.3em}
    \small{\raggedright Note. Network size is represented by the number of nodes. Each dot represents one run of related competing algorithm while lines representing linear approximation. Shaded area under lines representing 95\% confidence intervals.\par}
\end{figure}

Moreover, we estimated the growth rate of runtime as a function of network size (Fig.~\ref{fig02}). Although the range of LN sizes available from the snapshots does not allow for a precise empirical estimation of the time complexity class, the trend in Fig.~\ref{fig02} is well approximated by a linear function.

\section{Discussion}

BMSSP represents an important recent theoretical development in shortest-path algorithms and introduces a substantially different approach from traditional priority-queue-based methods.

While our results show that BMSSP does not outperform Dijkstra's algorithm in the evaluated setting, this outcome should not be interpreted as a contradiction of the theoretical guarantees. Rather, it highlights the well-known gap between asymptotic complexity and practical performance, particularly in systems where constant factors, memory behaviour, and implementation details dominate runtime.

Although BMSSP was slower in all evaluated scenarios, our experiments were not designed to identify the precise causes of the observed runtime differences. Several plausible explanations include memory-access patterns, recursion overhead, auxiliary data-structure management, and implementation-level constant factors. However, because we did not collect profiling metrics such as cache-miss counts, memory-allocation statistics, or recursion-depth measurements, these explanations should be viewed as hypotheses rather than established causal mechanisms.

Closely related to this is the role of control flow and recursion overhead. BMSSP's recursive structure introduces repeated function calls, stack frame management, and branching logic that are largely absent in a typical Dijkstra implementation. Modern compilers and processors are highly optimized for predictable, iterative workloads, whereas deeply recursive algorithms can incur penalties due to branch misprediction and stack management overhead. In our implementation, these factors appear to contribute substantially to the observed runtime differences.

Another important consideration is the constant-factor overhead inherent in BMSSP's design. While BMSSP reduces the need for global sorting, it replaces this with multiple localized relaxations and partitioning steps. Each of these operations introduces additional computational overhead that may dominate runtime for graph sizes typical of current Lightning Network snapshots. In other words, the asymptotic improvement may only become practically relevant at significantly larger scales than those currently observed in real-world PCNs.

Our findings also suggest that algorithm engineering plays a crucial role in determining practical performance. Dijkstra's algorithm has benefited from decades of optimization, including highly tuned priority queue implementations, memory-efficient data layouts, and hardware-aware optimizations. In contrast, BMSSP is a very recent development, and existing implementations—including ours—prioritize correctness and fidelity to the theoretical model over low-level performance tuning. This imbalance naturally favours modern Dijkstra's implementation in empirical comparisons.

It is also important to consider the interaction between algorithm design and graph structure. Although the Lightning Network is sparse, it exhibits additional structural properties such as skewed degree distributions and the presence of hub nodes. These characteristics may implicitly favour Dijkstra's exploration strategy, which incrementally expands a frontier guided by a priority queue. BMSSP's partition-based approach, while theoretically efficient, may not yet exploit these structural regularities effectively in its current form.

Finally, our evaluation raises broader questions about the conditions under which BMSSP might achieve practical advantages. It is plausible that BMSSP would perform better on significantly larger graphs, on architectures optimized for parallel or distributed execution, or with carefully engineered data structures that minimize recursion and memory overhead. Additionally, hybrid approaches that combine elements of Dijkstra’s priority-based exploration with BMSSP’s partitioning strategy may offer a promising direction for future work.

We have done our best to ensure a faithful implementation of the BMSSP algorithm over a performant one, and we are confident that with enough time, implementations of BMSSP will continue to improve in speed and capability. After all, the first implementation of Dijkstra's ran in $O(n^2)$ time and was not improved until 1977 when Donald Johnson introduced the priority queue as a practical speedup\cite{Johnson1977}.

In summary, our results do not diminish the theoretical significance of BMSSP, but rather emphasize the importance of bridging the gap between theory and practice. Achieving practical speedups will likely require not only improved implementations, but also a deeper understanding of how modern hardware and real-world graph structures interact with advanced shortest path algorithms.

We view this work as an initial step toward addressing the limitations of BMSSP on real-world problems. Many challenges remain ahead, particularly speeding up BMSSP using better techniques for modern computing architectures, as well as testing BMSSP against Dijkstra's variants on a number of other real-world networks. Thus, we invite the broader research community to collaborate in exploring these directions and help us build on this foundation moving forward.

Future work should incorporate profiling tools (e.g., hardware performance counters, cache-miss statistics, allocation tracing, and call-stack analysis) to determine which implementation characteristics most strongly influence practical performance.

\section{Conclusion}

This paper presents the first evaluation of a BMSSP implementation in the context of payment channel networks, specifically the Lightning Network. Building on the theoretical breakthrough of Duan et al., we implemented the BMSSP algorithm and compared it with Dijkstra’s algorithm using real-world Lightning Network topology data. 

Across all evaluated Lightning Network snapshots, BMSSP exhibited higher runtimes than Dijkstra's algorithm despite its stronger asymptotic guarantees. These findings suggest that current BMSSP implementations do not yet provide practical advantages on LN graphs of contemporary size. The primary contribution of this work is therefore not the demonstration of a faster routing algorithm, but rather an LN-specific empirical assessment of a recent theoretical breakthrough. The results highlight the importance of validating asymptotic improvements through implementation and systems-level evaluation before drawing conclusions about practical deployment.

To the best of our knowledge, this work represents the first empirical evaluation of BMSSP on Lightning Network topologies \cite{valko2025outperforming}. Our implementation builds upon the previously reported Rust prototype \cite{Paranjpe2025} and extends its application to payment-channel-network routing. The study complements subsequent implementations developed in C++ \cite{castro2025} and Python \cite{makowski2025bmsspy} by providing domain-specific evidence from real LN graph snapshots.

\section*{Data and code availability}
All data, developed components and algorithms are referenced or can be found in public repositories: \url{https://github.com/rap2363/ssps} and \url{https://github.com/ellariel/ln-bmssp-evaluation}

\bibliographystyle{splncs04}
\bibliography{paper}

@article{datapaper2025,
author = {Danila Valko and Jorge {Marx G\'omez}},
journal = {Scientific Data},
title = {Geolocated lightning network topology snapshots: A dataset covering 2019–2023},
year = {2025},
doi = {10.1038/s41597-025-06413-7},
}

@inproceedings{mpp1,
author = {Rahimpour, Sonbol and Khabbazian, Majid},
title = {Spear: fast multi-path payment with redundancy},
year = {2021},
isbn = {9781450390828},
publisher = {Association for Computing Machinery},
address = {New York, NY, USA},
doi = {10.1145/3479722.3480997},
booktitle = {Proceedings of the 3rd ACM Conference on Advances in Financial Technologies},
pages = {183–191},
numpages = {9},
location = {Arlington, Virginia},
series = {AFT '21}
}

@misc{mpp2,
      title={Optimally Reliable \& Cheap Payment Flows on the Lightning Network}, 
      author={Rene Pickhardt and Stefan Richter},
      year={2021},
      eprint={2107.05322},
      archivePrefix={arXiv},
      primaryClass={cs.NI},
      doi  = {10.48550/arXiv.2107.05322},
}

@misc{makowski2025bmsspy,
  author       = {Makowski, Connor and Guter, Willem and Russell, Timothy and Saragih, Austin and Castro, Lucas},
  title        = {BMSSPy: A Python Package and Empirical Comparison of Bounded Multi-Source Shortest Path Algorithm},
  archivePrefix={SSRN},
  year         = {2025},
  month        = {11},
  doi          = {10.2139/ssrn.5777186},
}

@misc{castro2025,
      title={Implementation and Brief Experimental Analysis of the Duan et al. (2025) Algorithm for Single-Source Shortest Paths}, 
      author={Lucas Castro and Thailsson Clementino and Rosiane de Freitas},
      year={2025},
      month        = {11},
      eprint={2511.03007},
      archivePrefix={arXiv},
      primaryClass={cs.DS},
      doi = {10.48550/arXiv.2511.03007},
}

@misc{valko2025outperforming,
      title={Outperforming Dijkstra on Sparse Graphs: The Lightning Network Use Case}, 
      author={Danila Valko and Rohan Paranjpe and Jorge Marx Gómez},
      year={2025},
      month        = {09},
      eprint={2509.13448},
      archivePrefix={arXiv},
      primaryClass={cs.PF},
      doi = {10.48550/arXiv.2509.13448},
}

@inproceedings{Duan2025,
author = {Duan, Ran and Mao, Jiayi and Mao, Xiao and Shu, Xinkai and Yin, Longhui},
title = {Breaking the Sorting Barrier for Directed Single-Source Shortest Paths},
year = {2025},
isbn = {9798400715105},
publisher = {Association for Computing Machinery},
address = {New York, NY, USA},
doi = {10.1145/3717823.3718179},
booktitle = {Proceedings of the 57th Annual ACM Symposium on Theory of Computing},
pages = {36–44},
numpages = {9},
location = {Prague, Czechia},
series = {STOC '25}
}

@misc{Duan2025arXiv,
      title={Breaking the Sorting Barrier for Directed Single-Source Shortest Paths}, 
      author={Ran Duan and Jiayi Mao and Xiao Mao and Xinkai Shu and Longhui Yin},
      year={2025},
      eprint={2504.17033},
      archivePrefix={arXiv},
      primaryClass={cs.DS},
      doi = {10.48550/arXiv.2504.17033},
}

@article{Zabka2024,
title = {A centrality analysis of the Lightning Network},
journal = {Telecommunications Policy},
volume = {48},
number = {2},
pages = {102696},
year = {2024},
issn = {0308-5961},
doi = {10.1016/j.telpol.2023.102696},
author = {Philipp Zabka and Klaus-T. Förster and Christian Decker and Stefan Schmid}
}

@INPROCEEDINGS{Rohrer2019,
author = { Rohrer, Elias and Malliaris, Julian and Tschorsch, Florian },
booktitle = { 2019 IEEE European Symposium on Security and Privacy Workshops (EuroS\&PW) },
title = {{ Discharged Payment Channels: Quantifying the Lightning Network's Resilience to Topology-Based Attacks }},
year = {2019},
volume = {},
ISSN = {},
pages = {347-356},
doi = {10.1109/EuroSPW.2019.00045},
publisher = {IEEE Computer Society},
address = {Los Alamitos, CA, USA},
month =Jun}

@article{Valko2024,
title = {Reducing CO2 emissions in a peer-to-peer distributed payment network: Does geography matter in the lightning network?},
journal = {Computer Networks},
volume = {243},
pages = {110297},
year = {2024},
issn = {1389-1286},
doi = {https://doi.org/10.1016/j.comnet.2024.110297},
author = {Danila Valko and Daniel Kudenko}
}

@misc{Paranjpe2025,
  author       = {Paranjpe, Rohan},
  title        = {Breaking the Shortest Path Barrier: A Deep Dive into BMSSP},
  year         = {2025},
note          = "Retrieved March 11, 2026 from \url{https://rohanparanjpe.substack.com/p/breaking-the-shortest-path-barrier}",
}

@misc{Saraswathi2025,
      title={An Exposition of Pathfinding Strategies Within Lightning Network Clients}, 
      author={Sindura Saraswathi and Christian Kümmerle},
      year={2025},
      eprint={2410.13784},
      archivePrefix={arXiv},
      primaryClass={cs.NI},
      doi = {10.48550/arXiv.2410.13784},
}

@article{Johnson1977,
author = {Johnson, Donald B.},
title = {Efficient Algorithms for Shortest Paths in Sparse Networks},
year = {1977},
issue_date = {Jan. 1977},
publisher = {Association for Computing Machinery},
address = {New York, NY, USA},
volume = {24},
number = {1},
issn = {0004-5411},
doi = {10.1145/321992.321993},
journal = {J. ACM},
month = jan,
pages = {1–13},
numpages = {13}
}

@misc{lngossip,
  title = {Lightning Network Research — Topology Datasets},
  author = {Decker, Christian},
  doi = {10.5281/zenodo.4088530},
  year = {2023},
note          = "Retrieved March 11, 2026 from \url{https://github.com/lnresearch/topology}",
}

@inproceedings{KumbleAnonimity,
author = {Kumble, Satwik Prabhu and Epema, Dick and Roos, Stefanie},
title = {How Lightning’s Routing Diminishes Its Anonymity},
year = {2021},
isbn = {9781450390514},
publisher = {Association for Computing Machinery},
address = {New York, NY, USA},
doi = {10.1145/3465481.3465761},
booktitle = {Proceedings of the 16th International Conference on Availability, Reliability and Security},
articleno = {13},
numpages = {10},
location = {Vienna, Austria},
series = {ARES 21}
}

@article{Thorup,
author = {Thorup, Mikkel},
title = {Undirected single-source shortest paths with positive integer weights in linear time},
year = {1999},
issue_date = {May 1999},
publisher = {Association for Computing Machinery},
address = {New York, NY, USA},
volume = {46},
number = {3},
issn = {0004-5411},
doi = {10.1145/316542.316548},
journal = {J. ACM},
month = may,
pages = {362–394},
numpages = {33}
}

\end{document}